\documentclass[aps,prl,twocolumn,superscriptaddress,showpacs,floatfix,10pt]{revtex4-2}

\usepackage{amsmath,amssymb,bm,graphicx,xcolor,braket}

\usepackage[backref=none,
bookmarksnumbered=true,
bookmarks=true,
bookmarksopen=true,
colorlinks=true,
citecolor=blue,
linkcolor=blue,
anchorcolor=green,
urlcolor=blue,unicode=false]{hyperref}

\newcommand{\bhat}{{\bf{b}}}
\newcommand{\lbold}{\boldsymbol{\ell}}
\newcommand{\lpar}{\ell_{\parallel}}
\newcommand{\lperp}{\ell_{\perp}}

\begin{document}

\title{How quantum is quantum geometry?}

\author{Ady Stern}
\affiliation{Department of Condensed Matter Physics, Weizmann Institute of Science, Rehovot 7610001, Israel}

\author{Felix von Oppen}
\affiliation{\mbox{Dahlem Center for Complex Quantum Systems, Fachbereich Physik, and Halle-Berlin-Regensburg}\\ Cluster of Excellence CCE, Freie Universit\"at Berlin, 14195 Berlin, Germany}

\begin{abstract}
The quantum geometric tensor -- the Berry curvature together with the quantum
metric -- now underlies a long list of observables, from the anomalous Hall effect
to the superfluid weight of a flat band. We ask which of these observables
actually require quantum mechanics. To answer this question, we study a purely classical
system: a point particle carrying a classical magnetic moment $\lbold$ that
precesses in a momentum-dependent magnetic field $\mathbf{B}(\mathbf{p})$. Within
Hamiltonian classical mechanics, the component of $\lbold$ along the field
reproduces the Berry-curvature phenomena, while its precessing transverse
component reproduces the quantum-metric phenomena. The particle acquires a
position spread whose second moment is the metric, an orbital magnetic moment,
and -- most strikingly -- an inertial mass generated by a position-dependent force, and with it a nonzero Drude weight in a system that is nominally dispersionless. 
\end{abstract}

\maketitle

\textit{Introduction}.---Over the past decade, quantum geometry has joined the energy dispersion as an essential characteristic of Bloch bands ~\cite{Torma2023review,Liu2024NSR,Yu2025review}. The geometry of a band's Bloch states is encoded in the quantum geometric tensor (QGT), which splits into a real symmetric part -- the Fubini-Study metric $g_{ij}$ -- and an imaginary antisymmetric part proportional to the Berry curvature $\Omega_{ij}$~\cite{ProvostVallee1980,Berry1984}.  The Berry curvature has long been understood to control the anomalous velocity of Bloch electrons and the intrinsic anomalous Hall effect~\cite{Sundaram1999,Xiao2010,Nagaosa2010AHE}. More recently the quantum metric has been shown to be equally consequential. It sets the superfluid weight of flat-band superconductors~\cite{Peotta2015,Torma2022NRP,Iskin2018}, the intrinsic nonlinear Hall effect~\cite{Gao2023NLH,Wang2023NLH}, resonant optical responses~\cite{Ahn2022Riemannian}, and the strength of electron--phonon coupling~\cite{Yu2024EPC}, and has now been measured~\cite{Kang2024measure}. The two parts are tied together by the positivity of the tensor. 

The quantum geometric tensor encodes how Bloch states within a band evolve in Hilbert space as momentum is varied. It is defined by
\begin{equation}
  \chi_{ij}(\mathbf{k}) = \big\langle \partial_{k_i} u_{\mathbf{k}} \big| 1 - P_{\mathbf{k}} \big|
  \partial_{k_j} u_{\mathbf{k}} \big\rangle = g_{ij} - \tfrac{i}{2}\,\Omega_{ij},
  \label{eq:qgt}
\end{equation}
where $|u_{\bf{k}}\rangle$ is the periodic part of the Bloch state and $P_{\bf{k}}=|u_{\bf{k}}\rangle\langle u_{\bf{k}}|$ projects onto it. 
Notice that energies do not appear in Eq.~(\ref{eq:qgt}) at all. 

In this work we examine the extent to which quantum geometry is indeed quantum. We introduce a classical system, and show that its response manifests most of the properties associated with quantum geometry. 

Our system is a particle carrying a classical magnetic moment in a momentum-dependent magnetic field. We derive the coupled equations of motion for the magnetic moment and for the motion of the particle, and show that most of the phenomena listed above originate from the effect of the fast dynamics of the magnetic moment on the slow dynamics of the particle. The classical counterpart of the Berry phase in systems composed of slow and fast degrees of freedom, and its effect on the classical mechanics of the slow degrees of freedom, were studied before \cite{Aharonov1992,Hannay1985,Dalessio2014,Kolodrubetz2017,Barrera2026}. Here, we focus on coupling the fast and slow degrees of freedom through the momentum, and on the classical analogs of the quantum metric.

A classical system is defined by its dynamical degrees of freedom and its Hamiltonian, from which equations of motion are derived. The solution of these equations fully describes the system's dynamics. The dynamical degrees of freedom in the system we consider are the magnetic moment $\boldsymbol{\ell} = (\ell_{x},\ell_{y},\ell_{z})$, and the particle's position $\mathbf{r}$ and momentum $\bf{p}$, whose number of components depends on the dimension. Position and momentum are canonically conjugate with Poisson brackets $\{r_i,p_j\}=\delta_{ij}$. The moment obeys the angular-momentum algebra $\{\ell_i,\ell_j\}=\epsilon_{ijk}\ell_k$, so that $|\lbold|$ is a constant of motion and its phase space is the sphere $S^2$. Nothing here is quantized: $\lbold$ is a classical vector of arbitrary length. 

We take the Hamiltonian $H=H(\mathbf{r},\mathbf{p},\lbold)$ to be
\begin{equation}
  H = - \lbold \cdot \mathbf{B}(\mathbf{p}) + V(\mathbf{r}).
  \label{eq:H}
\end{equation}
The particle is subject to a scalar potential and couples to the moment through a magnetic field which depends on the particle's momentum. We do not include a kinetic energy $\mathbf{p}^2/2m$, having in mind the analogy to electrons in a perfectly flat band with infinite effective mass. The equations of motion take the form
\begin{equation}
  \dot{\lbold} = \lbold\times\mathbf{B}, \qquad
  \dot{\mathbf{p}} = -\boldsymbol{\nabla}V(\mathbf{r}), \qquad
  \dot{r}_i = - \lbold\cdot\frac{\partial \mathbf{B}}{\partial p_i}.
  \label{eq:eom}
\end{equation}
Generally, the force $\mathbf{F} = -\boldsymbol{\nabla}V(\mathbf{r})$ changes the momentum, this changes the field $\mathbf{B}(\mathbf{p})$, and the response of the
moment to a field whose direction changes in time is where all of the geometry resides. The last equation in Eq.~(\ref{eq:eom}) converts that response into motion. 

We now turn to the dynamics of $\lbold$. 
Two useful choices sharpen the discussion. First, we write
$\mathbf{B}(\mathbf{p}) = B\,\bhat(\mathbf{p})$, with $\bhat$ being a unit vector, and take the magnitude $B$ to be independent
of $\mathbf{p}$. Thus, only the \emph{direction} of the field varies, and we can draw again an analogy to the quantum problem of a perfectly flat band. Second, we take $B$ to be the large scale in the problem and retain only those effects that do \emph{not} vanish as $B\to\infty$. In this limit the precession is fast compared to everything else, $\lpar = \lbold\cdot\bhat$ is an adiabatic invariant and becomes the key quantity.

When $\mathbf{b}$ is time independent, the moment's dynamics is elementary. It precesses
about the field,
\begin{equation}
  \lbold(t) = \lpar\,\bhat
  + \lperp\big({\mathbf{e}}_1\cos Bt - {\mathbf{e}}_2\sin Bt\big),
  \label{eq:precession}
\end{equation}
where the unit vectors ${\mathbf{e}}_{1,2}$ span the plane perpendicular to $\bhat$ with
${\mathbf{e}}_1\times{\mathbf{e}}_2=\bhat$. The two components $\lpar$ and $\lperp$ with $\lpar^2+\lperp^2=|\lbold|^2$ organize everything
that follows. We will find that $\lpar$ generates the anomalous velocity and $\lperp$ the quantum-metric phenomenology.

We start by setting $\lperp=0$ and letting $\bhat$ vary slowly in time, as it will once the particle's momentum becomes time dependent by the application of a force $\mathbf{F}$. To leading order in $B$, the magnetic moment follows the magnetic field, such that $\boldsymbol{\ell}\parallel \mathbf{b}$ at all times. Deviations from this relation are of order $1/B$. While small, these deviations give contributions of order $B^0$ to the velocity once substituted into the equation of motion. The tilt of order $1/B$ relative to $\boldsymbol{\ell}\parallel \mathbf{b}$ is obtained by writing the equation of motion for $\boldsymbol{\ell}$ in a moving reference frame \cite{Aharonov1992} (see End Matter for details). One parallel transports the original reference frame such that the instantaneous $\mathbf{b}$ is rotated to the $\mathbf{z}$-axis, and the parallel-transport axis of rotation is orthogonal to both $\mathbf{b}$ and $\mathbf{\dot b}$. Denoting vectors in the moving reference frame by a prime, the rotated magnetic moment $\boldsymbol{\ell}'$ satisfies 
\begin{equation}
  \frac{d\lbold'}{dt} = \lbold'\times \mathbf{B}_\mathrm{eff}' , \qquad
  \mathbf{B}_\mathrm{eff}' = B\,\mathbf{z} + (\bhat \times \dot{\bhat})'.
  \label{eq:Bprime}
\end{equation}
The moment $\boldsymbol{\ell}$ no longer precesses about $B\bhat$ but about a slightly tilted effective
field $\mathbf{B}_\mathrm{eff}$. The adiabatic limit is the statement that the tilt is small,
i.e.\ that $|\dot{\bhat}|/B \ll 1$. Including the correction to the naive $\boldsymbol{\ell}\parallel \mathbf{b}$ assumption gives 
\begin{equation}
  \lbold(t) = \lpar\Big(\bhat + \frac{1}{B}\,\bhat\times\dot{\bhat}\Big)
  + \mathcal{O}(B^{-2}).
  \label{eq:refined}
\end{equation}
The correction is perpendicular to both $\bhat$ and $\dot{\bhat}$. 

Equation (\ref{eq:refined}) is consistent with the adiabatic theorem, which
asserts the action is an invariant of the adiabatic evolution. The action-angle variables for the Hamiltonian
$\lbold\cdot\mathbf{B}$   of the moment are the projection $\lpar=\lbold\cdot\bhat$ and the precession angle about $\bhat$. As
\begin{equation}
  \frac{d}{dt}\big(\lbold\cdot\bhat\big)
  = \big(\lbold\times\mathbf{B}\big)\cdot\bhat + \lbold\cdot\dot{\bhat}
  = \lbold\cdot\dot{\bhat},
\end{equation}
conservation of the action requires only that $\lbold$ has no component along $\dot{\bhat}$. It does not preclude a component along $\bhat\times\dot{\bhat}$, consistent with  Eq.~(\ref{eq:refined}). The adiabatic theorem permits the tilt; it is
the geometry that fixes it. 

The tilt of $\boldsymbol{\ell}$ of order $1/B$ translates into a velocity of order $B^0$, the anomalous velocity.  Inserting $\lbold=\lpar\bhat$ into the velocity in Eq.~(\ref{eq:eom}) gives a vanishing velocity, since $\bhat\cdot\partial_i\bhat = 0$. However, the correction in Eq.~(\ref{eq:refined}) does endow the particle with a velocity. Using $\dot{\bhat} = (\partial\bhat/\partial p_j) \dot p_j$ and
$\mathbf{F}=\dot{\mathbf{p}}$, we find
\begin{equation}
  \dot{r}_i
  = \lpar\;\bhat\cdot\Big(\frac{\partial\bhat}{\partial p_i}\times
  \frac{\partial\bhat}{\partial p_j}\Big) F_j .
  \label{eq:anomalous}
\end{equation}
This  velocity is transverse to the applied force, proportional to it, and independent of $B$. It vanishes identically in one dimension, but is generically nonzero in higher dimensions. We note that the force $\bf F$ that produces it does not accelerate the particle -- a consequence of $B$ being independent of momentum. 

Equation~(\ref{eq:anomalous}) is the classical anomalous velocity, and the comparison with the quantum expression is evident. For a quantum two-level system in the field $\mathbf{B}(\mathbf{p})$, the eigenstate aligned with $\bhat$ has Berry curvature $\Omega_{ij} = \epsilon_{ij} \Omega = \tfrac12\,\bhat\cdot(\partial_i\bhat\times\partial_j\bhat)$ in momentum space, so that Eq.~(\ref{eq:anomalous}) reads $\dot{r}_i =  2\lpar\,\Omega_{ij}F_j$. Setting $\lpar=\hbar/2$ reproduces the standard result $\dot{\mathbf{r}} = \mathbf{F} \times \boldsymbol{\Omega} $ of Refs.~\cite{Karplus1954,Xiao2010}. Classically $\lpar$ is a continuous parameter, while it is quantized in quantum mechanics. This analysis is close in spirit to the observation of Ref.~\cite{Aharonov1992} that the forces accompanying Berry's phase when the fast and slow degrees of freedom are coupled through the particle's position are classical.

We now turn on $\lperp$, which was set to zero so far. As we show below, $\lperp$ leads to dynamical phenomena that are commonly associated with the quantum metric. 

First we note that even when $\mathbf{p}$ is a constant of motion, the velocity in
Eq.~(\ref{eq:eom}) no longer vanishes with the precessing solution in Eq.\ (\ref{eq:precession}). Instead, it oscillates, with a short time period $2\pi/B$ and a large amplitude $\lperp B$,
\begin{equation}
  \delta\dot{r}_i = - B\lperp
  \big({\mathbf{ e}}_1\cos Bt - {\mathbf{ e}}_2\sin Bt\big)\cdot
  \frac{\partial\bhat}{\partial p_i}  .
  \label{eq:oscvel}
\end{equation}
The displacement amplitude, then, is independent of $B$,
\begin{equation}
  \xi_i = \lperp\,\left|\frac{\partial \mathbf{b}}{\partial p_i}\right| .
  \label{eq:xi}
\end{equation}
The extent of the motion is  determined entirely by how fast the field direction turns with momentum. It is, in this sense, geometric.

Even in the absence of a driving force, the oscillatory motion of the particle has two consequences. The first is a spread of the particle's position. Averaging the excursion over a precession period gives
\begin{equation}
  \big\langle \delta r_i\, \delta r_j \big\rangle
  = \frac{\lperp^2}{2}\,
  \frac{\partial\bhat}{\partial p_i}\cdot\frac{\partial\bhat}{\partial p_j}.
  \label{eq:spread}
\end{equation}
The particle is not at a point -- its position is  smeared over a region whose second moment {is} geometric. This is the classical origin of the bound imposed by the quantum metric on  the spread of
a Wannier function \cite{Marzari1997,Marzari2012}. 

The second observable is a magnetic moment. The oscillation of Eq.~(\ref{eq:oscvel}) is a closed orbit traversed at frequency $B$. In more than one spatial dimension these oscillations may create a current loop and, if the particle is charged, an orbital moment. Evaluating $\mathbf{m} = \tfrac{e}{2}\langle \delta\mathbf{r}\times\delta\dot{\mathbf{r}}\rangle$ in two dimensions, with the average taken over a period of the motion, gives
\begin{equation}
  m_z = \tfrac{e}{2}\,B\lperp^2\;
  \bhat\cdot\Big(\frac{\partial\bhat}{\partial p_x}\times
  \frac{\partial\bhat}{\partial p_y}\Big)
  = e B \lperp^2\,\Omega_{xy}.
  \label{eq:moment}
\end{equation}
Unlike the anomalous velocity, the magnetic moment is of order $B$. It is proportional to the classical ``Berry curvature". This structure is reminiscent of the wave-packet orbital moment of a model of two perfectly flat bands \cite{ChangNiu2008,Thonhauser2005}. Notice that unlike the anomalous velocity, this quantity requires $\ell_\perp\ne 0$.

Equations (\ref{eq:anomalous}), (\ref{eq:spread}), and (\ref{eq:moment}) motivate the definition of the classical analog of the quantum geometric tensor,
\begin{equation}
    \chi_{ij}=\ell^2\left(\frac{\partial\bhat}{\partial p_i}\cdot\frac{\partial\bhat}{\partial p_j}-i\frac{\partial{\mathbf b}}{\partial p_i}
  \times\frac{\partial{\mathbf b}}{\partial p_j}\cdot{\mathbf b}\right).
  \label{eq:classical}
\end{equation}
Similar to its quantum counterpart, this classical $\chi_{ij} = \ell^2(g_{ij} - 2i \Omega_{ij})$ satisfies $\mathrm{tr} g \geq 4 |\Omega_{xy}|$. 

So far, the fast oscillations produced a spread of the position of the particle, and a magnetic moment, but no net transport. The velocity in Eq.~(\ref{eq:oscvel}) averages to zero. We now turn to show how these oscillations translate to  motion. 

As seen in the equation of motion for the velocity, the oscillations in $\boldsymbol{\ell}$ can translate to a velocity which does not average to zero, when  $\mathbf{p}$ and consequently $\partial\bhat/\partial p_i$ oscillate. This requires  a force which depends on position. A uniform force changes the momentum at a constant rate, making $\partial\bhat/\partial{p}_i$ drift rather than oscillate, and leaving 
the average velocity zero. The absence of a $\mathbf{p}^2/2m$ term in the Hamiltonian (\ref{eq:H}) implies the absence of acceleration for a uniform force.

The effect of a position-dependent force is most simply exemplified if we consider the force exerted by a spring, $\mathbf{F}=-k\mathbf{r}$. The
oscillation of the position feeds back into the momentum,
\begin{equation}
  \delta p_i(t) = -k\!\int\! \delta r_i\,dt
  \;\simeq\; - \frac{k\lperp}{B}\,\frac{\partial\bhat}{\partial{p}_i} \cdot
  \big({\mathbf{e}}_1\cos Bt - {\mathbf{e}}_2\sin Bt\big),
\end{equation}
and this in turn modulates the coupling that produced the oscillation in the
first place,
\begin{equation}
  \delta \left(\frac{\partial\bhat}{\partial p_i}\right)
  \simeq  \frac{\partial^2\bhat}{\partial p_i\partial p_j}\,\delta p_j . 
  \label{eq:posc}
\end{equation}
Inserting Eq.~\eqref{eq:posc} into Eq.~(\ref{eq:oscvel}) gives the product of two quantities oscillating at the same frequency, which does not vanish when averaged. The two factors involving $B$ -- one from the velocity amplitude, one from
the $1/B$ in $\delta\mathbf{p}$ -- cancel, and the resulting time-averaged velocity,
\begin{equation}
  \langle \dot{r}_i \rangle =
  \frac{k\lperp^2}{4}\,\frac{\partial}{\partial p_i}
  \left(\frac{\partial\bhat}{\partial {p}_j}\cdot \frac{\partial\bhat}{\partial {p}_j}\right) 
  \label{eq:drift}
\end{equation}
is $B$-independent.

Equation~(\ref{eq:drift}) deserves a few comments. The right-hand side is a momentum derivative, so the dependence of the velocity on momentum is what one would obtain from an effective kinetic energy
\begin{equation}
  H_{\rm kin}(\mathbf{p}) =
  \frac{k\lperp^2}{4} \frac{\partial\bhat}{\partial {p}_j}\cdot \frac{\partial\bhat}{\partial {p}_j}. 
  \label{eq:Hkin}
\end{equation}
A dispersion has been generated in the effective Hamiltonian for the slow degrees of freedom, i.e., the particle's position and momentum. The induced momentum dependence is geometric and given by the trace of the metric tensor. Even for a harmonic potential which depends only on one coordinate, say $\frac{1}{2}kx^2$, the energy will in general depend on all momentum components, as determined by  the derivatives $\frac{\partial\bhat}{\partial p_j}$. Expanding about a minimum gives an effective
mass
\begin{equation}
  m^* \propto \big(k\lperp^2\big)^{-1}.
\end{equation}
It diverges as the force becomes uniform ($k\to0$) or as the precession is switched off ($\lperp\to0$), but is  otherwise finite. We emphasize the character of the result: The particle acquires inertia from an interplay of a geometric oscillation with an inhomogeneous force.
The mass is, in the language of Eq.~(\ref{eq:xi}), the result of the particle experiencing different forces at  different points across the extent of the orbit.

Once the energy disperses with momentum, the application of a uniform force leads to acceleration, as it does for all massive particles.

We now show that two particles that mutually interact may endow one another with a mass. Guided by the above discussion, we first couple the particles, labeled $a = 1,2$, through a spring.  The joint Hamiltonian is 
\begin{equation}
H = - {\boldsymbol{\ell}_1\cdot {\bf b}({\bf p}_1}) - {\boldsymbol{\ell}_2\cdot{\bf b}({\bf p}_2)}+\frac{k}{2}({\bf r}_1-{\bf r_2})^2.
\label{twoparticles}
\end{equation}
The total momentum ${\bf P} = {\bf p}_1+{\bf p}_2$ is conserved, motivating us to write ${\bf p}_{1,2} = {\bf P}\pm \Delta {\bf p}/2$.
For $\Delta{\bf p} = {\bf p}_1 - {\bf p}_2$ to oscillate at the precession frequency, we need ${\bf r}_1-{\bf r}_2$ to oscillate. While both ${\bf r}_1$ and ${\bf r}_2$ oscillate due to the precession, their difference does not have to. However, small variations of the frequency $B$ with momentum, or of $\ell_{1,\perp},\ell_{2,\perp}$, remove all correlations between the oscillations of ${\bf r}_1$ and ${\bf r}_2$ and induce an effective kinetic energy
\begin{equation}
  H_{\rm kin}= \frac{k}{4} \sum_{a = 1,2}
  \ell_{a,\perp}^2 \,\frac{\partial\bhat}{\partial p_{a,j}} \cdot \frac{\partial\bhat}{\partial p_{a,j}}.
  \label{eq:twoparticle}
\end{equation}
Equation (\ref{eq:twoparticle}) depends on $\mathbf{p}_1$ and $\mathbf{p}_2$
separately, and therefore on both the center-of-mass and relative momenta $\mathbf{P}$ and $\Delta\mathbf{p}$.

When a uniform force acts on the two particles, the center-of-mass dynamical momentum varies linearly in time ${\bf P}\rightarrow {\bf P}+{\bf F}t$ and for short times the two particles accelerate at a rate proportional to the force $F$. 
Their effective mass tensor becomes 
\begin{equation}
  \left (\frac{1}{m}\right)_{ij}\;\sim\; 
  \frac{\partial^2}{\partial p_i\partial p_j}\, H_{\rm kin}
\end{equation}
with $H_{\rm kin}$ given by Eq. (\ref{eq:twoparticle}). 
The attractive interaction between the two particles endows their average position with an effective mass, which does not exist for each of the particles. Quantum mechanically, when the two particles are electrons, they form a Cooper pair. When many such pairs are present, they may condense into a super-conductor \cite{Peotta2015,Torma2022NRP,Resta2017Drude,Shinada2025bounds}. 

Our choice of a harmonic force was driven by mathematical convenience. We now generalize the analysis to forces  between the particles with arbitrary distance dependence. The essential physics can be illustrated in one dimension. The time-averaged velocities of the particles simplify to \begin{equation}
  \dot{x}_a = - B\lperp \frac{\partial^2\bhat}{\partial{p}_a^2} \cdot
  \langle \left({\mathbf{e}}_1\cos Bt - {\mathbf{e}}_2\sin Bt\right) {p} \rangle ,
\end{equation}
and ${\dot p}_a = -\partial_a V(x_1 - x_2)$. Decomposing the potential into Fourier components $V_q$ and averaging $e^{iq(x_1 - x_2)}$ over the fast oscillations produces Bessel-function form factors (see End Matter for details),
\begin{equation}
  {\dot x}_1 =   \partial_{p_1}
  \int \frac{dq}{2\pi}  V_q e^{iq(x_1 - x_2)} 
 J_0 (q\xi_1)
  J_0(q\xi_2).
\end{equation}
Here, the arguments of the Bessel functions contain the displacement amplitudes $\xi_a$ defined in Eq.\ \eqref{eq:xi}. We assumed again that the precession frequencies of the two particles are not identical. Expanding the Bessel functions to leading nonvanishing oder in $q$ recovers Eq.~(\ref{eq:twoparticle}) with a local ``spring constant"  $V''(x_1-x_2)$.

The particle velocities can be written as a momentum derivative of an emergent kinetic-energy contribution to the Hamiltonian of the slow motion (see End Matter), 
\begin{equation}
H_\mathrm{kin} = \int \frac{dq}{2\pi}  V_q e^{iq(x_1 - x_2 )} 
 J_0 (q\ell_\perp \xi_1)
  J_0 (q\ell_\perp \xi_2).
\label{eq:Hkin_ibp}
\end{equation}
This expression is just the interaction potential of the two particles, averaged over the fast oscillations
\begin{equation}
    H_\mathrm{kin} = \langle V(x_1 - x_2) \rangle. 
    \label{eq:HkinavV}
\end{equation}
While we derived Eq.\ \eqref{eq:HkinavV} for one dimension, it remains valid for any spatial dimension. The momentum dependence enters through the displacement amplitudes $\xi_a = \xi_a(\mathbf{p}_a)$. This shows that as a result of the averaging over the fast oscillations, the two particles develop a joint inertial mass. Unlike a conventional kinetic energy, $H_\mathrm{kin}$ and with it the emergent mass, depend on the particle distance in addition to the momenta. In particular, for a finite-range interaction there is no inertia for distances beyond the interaction range, which is the range of the potential augmented by $\xi_1+\xi_2$. 

Our derivations,  from Eq.~(\ref{eq:H}) to Eq.~(\ref{eq:Hkin_ibp}), are based on classical
Hamiltonian mechanics. Remarkably, they showed that classical particles demonstrate the anomalous velocity, the
orbital magnetic moment, a position spread, and the
generation of an inertial mass by the interaction of two particles whose individual inertial masses are infinite. These observables are ordinarily attributed to quantum geometry.

When comparing the classical analogs we derived for the Berry curvature and the quantum metric to their quantum counterparts, the quantum derivative of an eigenstate with respect to momentum $|\partial_{\bf p}n(\bf p)\rangle$ is replaced by the derivative $\partial_{\bf p}{\bf b}({\bf p)}$. The $\hbar$ factor that appears in the quantum quantities is replaced by $\ell_\parallel$ for the anomalous velocity, and by $\lperp$ for the other observables we discussed. If we consider the Hamiltonian (\ref{eq:H}) as quantum mechanical, and consider the properties of its eigenstates in the classical limit ${\lbold}^2\rightarrow\infty$ and $\hbar\rightarrow 0$, the value of $\ell_\parallel$ is quantized to $\hbar$ times an integer between $-\ell$ and $\ell$. There is always a precessing component, since $\ell(\ell+1)-\ell^2 > 0$, but the phase of the precession is undefined, due to the uncertainty relation. 

Two consequences of quantum geometry are inherently quantum mechanical. The first is the magnetization as a thermodynamic property. The single-particle orbital moment of Eq.~(\ref{eq:moment}) is classical, but by the Bohr--van Leeuwen theorem~\cite{ashcroft1976solid} the equilibrium magnetization is quantum mechanical. The second is superconductivity as a Bose condensate of pairs. Our two particles acquire a joint inertial mass and hence a Drude weight in the absence of scattering, but nothing in Eq.~(\ref{eq:twoparticle}) produces off-diagonal long-range order, phase rigidity, or a Meissner effect. The flat-band statement that the superfluid weight is bounded by the integrated metric thus splits into a classical part -- the inverse mass -- and a quantum part -- the condensation. Beyond these two consequences, quantum mechanics also introduces the periodic Brillouin-zone structure to momentum space, which underlies the topological consequences of quantum geometry. 

To summarize, our analysis shows that most phenomena associated with quantum geometry are in fact of a classical origin, and originate from the interplay between slow and fast degrees of freedom of a dynamical system.

\begin{acknowledgments}
We thank Maximilian Rieger for discussions. We acknowledge support from the Minerva Stiftung, the DFG through CRC/Transregio 183 (Project Grant No.\ 277101999) and the Center for Chiral Electronics (German Excellence Strategy EXC3112/1 - 533767171), the Israel Science Foundation ISF (Grant No 1914/24), as well as ISF Quantum Science and Technology (2074/19). 
\end{acknowledgments}


\begin{thebibliography}{29}%
\makeatletter
\providecommand \@ifxundefined [1]{%
 \@ifx{#1\undefined}
}%
\providecommand \@ifnum [1]{%
 \ifnum #1\expandafter \@firstoftwo
 \else \expandafter \@secondoftwo
 \fi
}%
\providecommand \@ifx [1]{%
 \ifx #1\expandafter \@firstoftwo
 \else \expandafter \@secondoftwo
 \fi
}%
\providecommand \natexlab [1]{#1}%
\providecommand \enquote  [1]{``#1''}%
\providecommand \bibnamefont  [1]{#1}%
\providecommand \bibfnamefont [1]{#1}%
\providecommand \citenamefont [1]{#1}%
\providecommand \href@noop [0]{\@secondoftwo}%
\providecommand \href [0]{\begingroup \@sanitize@url \@href}%
\providecommand \@href[1]{\@@startlink{#1}\@@href}%
\providecommand \@@href[1]{\endgroup#1\@@endlink}%
\providecommand \@sanitize@url [0]{\catcode `\\12\catcode `\$12\catcode
  `\&12\catcode `\#12\catcode `\^12\catcode `\_12\catcode `\%12\relax}%
\providecommand \@@startlink[1]{}%
\providecommand \@@endlink[0]{}%
\providecommand \url  [0]{\begingroup\@sanitize@url \@url }%
\providecommand \@url [1]{\endgroup\@href {#1}{\urlprefix }}%
\providecommand \urlprefix  [0]{URL }%
\providecommand \Eprint [0]{\href }%
\providecommand \doibase [0]{https://doi.org/}%
\providecommand \selectlanguage [0]{\@gobble}%
\providecommand \bibinfo  [0]{\@secondoftwo}%
\providecommand \bibfield  [0]{\@secondoftwo}%
\providecommand \translation [1]{[#1]}%
\providecommand \BibitemOpen [0]{}%
\providecommand \bibitemStop [0]{}%
\providecommand \bibitemNoStop [0]{.\EOS\space}%
\providecommand \EOS [0]{\spacefactor3000\relax}%
\providecommand \BibitemShut  [1]{\csname bibitem#1\endcsname}%
\let\auto@bib@innerbib\@empty
\bibitem [{\citenamefont {T\"orm\"a}(2023)}]{Torma2023review}%
  \BibitemOpen
  \bibfield  {author} {\bibinfo {author} {\bibfnamefont {P.}~\bibnamefont
  {T\"orm\"a}},\ }\bibfield  {title} {\bibinfo {title} {Essay: Where can
  quantum geometry lead us?},\ }\href
  {https://doi.org/10.1103/PhysRevLett.131.240001} {\bibfield  {journal}
  {\bibinfo  {journal} {Phys. Rev. Lett.}\ }\textbf {\bibinfo {volume} {131}},\
  \bibinfo {pages} {240001} (\bibinfo {year} {2023})}\BibitemShut {NoStop}%
\bibitem [{\citenamefont {Liu}\ \emph {et~al.}(2025)\citenamefont {Liu},
  \citenamefont {Qiang}, \citenamefont {Lu},\ and\ \citenamefont
  {Xie}}]{Liu2024NSR}%
  \BibitemOpen
  \bibfield  {author} {\bibinfo {author} {\bibfnamefont {T.}~\bibnamefont
  {Liu}}, \bibinfo {author} {\bibfnamefont {X.-B.}\ \bibnamefont {Qiang}},
  \bibinfo {author} {\bibfnamefont {H.-Z.}\ \bibnamefont {Lu}},\ and\ \bibinfo
  {author} {\bibfnamefont {X.~C.}\ \bibnamefont {Xie}},\ }\bibfield  {title}
  {\bibinfo {title} {Quantum geometry in condensed matter},\ }\href
  {https://doi.org/10.1093/nsr/nwae334} {\bibfield  {journal} {\bibinfo
  {journal} {National Science Review}\ }\textbf {\bibinfo {volume} {11}},\
  \bibinfo {pages} {nwae334} (\bibinfo {year} {2025})}\BibitemShut {NoStop}%
\bibitem [{\citenamefont {Yu}\ \emph {et~al.}(2025)\citenamefont {Yu},
  \citenamefont {Bernevig}, \citenamefont {Queiroz}, \citenamefont {Rossi},
  \citenamefont {T{\"o}rm{\"a}},\ and\ \citenamefont {Yang}}]{Yu2025review}%
  \BibitemOpen
  \bibfield  {author} {\bibinfo {author} {\bibfnamefont {J.}~\bibnamefont
  {Yu}}, \bibinfo {author} {\bibfnamefont {B.~A.}\ \bibnamefont {Bernevig}},
  \bibinfo {author} {\bibfnamefont {R.}~\bibnamefont {Queiroz}}, \bibinfo
  {author} {\bibfnamefont {E.}~\bibnamefont {Rossi}}, \bibinfo {author}
  {\bibfnamefont {P.}~\bibnamefont {T{\"o}rm{\"a}}},\ and\ \bibinfo {author}
  {\bibfnamefont {B.-J.}\ \bibnamefont {Yang}},\ }\bibfield  {title} {\bibinfo
  {title} {Quantum geometry in quantum materials},\ }\href
  {https://doi.org/10.1038/s41535-025-00801-3} {\bibfield  {journal} {\bibinfo
  {journal} {npj Quantum Materials}\ }\textbf {\bibinfo {volume} {10}},\
  \bibinfo {pages} {101} (\bibinfo {year} {2025})}\BibitemShut {NoStop}%
\bibitem [{\citenamefont {Provost}\ and\ \citenamefont
  {Vall\'ee}(1980)}]{ProvostVallee1980}%
  \BibitemOpen
  \bibfield  {author} {\bibinfo {author} {\bibfnamefont {J.~P.}\ \bibnamefont
  {Provost}}\ and\ \bibinfo {author} {\bibfnamefont {G.}~\bibnamefont
  {Vall\'ee}},\ }\bibfield  {title} {\bibinfo {title} {Riemannian structure on
  manifolds of quantum states},\ }\href {https://doi.org/10.1007/BF02193559}
  {\bibfield  {journal} {\bibinfo  {journal} {Commun. Math. Phys.}\ }\textbf
  {\bibinfo {volume} {76}},\ \bibinfo {pages} {289} (\bibinfo {year}
  {1980})}\BibitemShut {NoStop}%
\bibitem [{\citenamefont {Berry}(1984)}]{Berry1984}%
  \BibitemOpen
  \bibfield  {author} {\bibinfo {author} {\bibfnamefont {M.~V.}\ \bibnamefont
  {Berry}},\ }\bibfield  {title} {\bibinfo {title} {Quantal phase factors
  accompanying adiabatic changes},\ }\href
  {https://doi.org/10.1098/rspa.1984.0023} {\bibfield  {journal} {\bibinfo
  {journal} {Proc. R. Soc. London A}\ }\textbf {\bibinfo {volume} {392}},\
  \bibinfo {pages} {45} (\bibinfo {year} {1984})}\BibitemShut {NoStop}%
\bibitem [{\citenamefont {Sundaram}\ and\ \citenamefont
  {Niu}(1999)}]{Sundaram1999}%
  \BibitemOpen
  \bibfield  {author} {\bibinfo {author} {\bibfnamefont {G.}~\bibnamefont
  {Sundaram}}\ and\ \bibinfo {author} {\bibfnamefont {Q.}~\bibnamefont {Niu}},\
  }\bibfield  {title} {\bibinfo {title} {{Wave-packet dynamics in slowly
  perturbed crystals: Gradient corrections and Berry-phase effects}},\ }\href
  {https://doi.org/10.1103/PhysRevB.59.14915} {\bibfield  {journal} {\bibinfo
  {journal} {Phys. Rev. B}\ }\textbf {\bibinfo {volume} {59}},\ \bibinfo
  {pages} {14915} (\bibinfo {year} {1999})}\BibitemShut {NoStop}%
\bibitem [{\citenamefont {Xiao}\ \emph {et~al.}(2010)\citenamefont {Xiao},
  \citenamefont {Chang},\ and\ \citenamefont {Niu}}]{Xiao2010}%
  \BibitemOpen
  \bibfield  {author} {\bibinfo {author} {\bibfnamefont {D.}~\bibnamefont
  {Xiao}}, \bibinfo {author} {\bibfnamefont {M.-C.}\ \bibnamefont {Chang}},\
  and\ \bibinfo {author} {\bibfnamefont {Q.}~\bibnamefont {Niu}},\ }\bibfield
  {title} {\bibinfo {title} {Berry phase effects on electronic properties},\
  }\href {https://doi.org/10.1103/RevModPhys.82.1959} {\bibfield  {journal}
  {\bibinfo  {journal} {Rev. Mod. Phys.}\ }\textbf {\bibinfo {volume} {82}},\
  \bibinfo {pages} {1959} (\bibinfo {year} {2010})}\BibitemShut {NoStop}%
\bibitem [{\citenamefont {Nagaosa}\ \emph {et~al.}(2010)\citenamefont
  {Nagaosa}, \citenamefont {Sinova}, \citenamefont {Onoda}, \citenamefont
  {MacDonald},\ and\ \citenamefont {Ong}}]{Nagaosa2010AHE}%
  \BibitemOpen
  \bibfield  {author} {\bibinfo {author} {\bibfnamefont {N.}~\bibnamefont
  {Nagaosa}}, \bibinfo {author} {\bibfnamefont {J.}~\bibnamefont {Sinova}},
  \bibinfo {author} {\bibfnamefont {S.}~\bibnamefont {Onoda}}, \bibinfo
  {author} {\bibfnamefont {A.~H.}\ \bibnamefont {MacDonald}},\ and\ \bibinfo
  {author} {\bibfnamefont {N.~P.}\ \bibnamefont {Ong}},\ }\bibfield  {title}
  {\bibinfo {title} {{Anomalous Hall effect}},\ }\href
  {https://doi.org/10.1103/RevModPhys.82.1539} {\bibfield  {journal} {\bibinfo
  {journal} {Rev. Mod. Phys.}\ }\textbf {\bibinfo {volume} {82}},\ \bibinfo
  {pages} {1539} (\bibinfo {year} {2010})}\BibitemShut {NoStop}%
\bibitem [{\citenamefont {Peotta}\ and\ \citenamefont
  {T\"orm\"a}(2015)}]{Peotta2015}%
  \BibitemOpen
  \bibfield  {author} {\bibinfo {author} {\bibfnamefont {S.}~\bibnamefont
  {Peotta}}\ and\ \bibinfo {author} {\bibfnamefont {P.}~\bibnamefont
  {T\"orm\"a}},\ }\bibfield  {title} {\bibinfo {title} {Superfluidity in
  topologically nontrivial flat bands},\ }\href
  {https://doi.org/10.1038/ncomms9944} {\bibfield  {journal} {\bibinfo
  {journal} {Nat. Commun.}\ }\textbf {\bibinfo {volume} {6}},\ \bibinfo {pages}
  {8944} (\bibinfo {year} {2015})}\BibitemShut {NoStop}%
\bibitem [{\citenamefont {T\"orm\"a}\ \emph {et~al.}(2022)\citenamefont
  {T\"orm\"a}, \citenamefont {Peotta},\ and\ \citenamefont
  {Bernevig}}]{Torma2022NRP}%
  \BibitemOpen
  \bibfield  {author} {\bibinfo {author} {\bibfnamefont {P.}~\bibnamefont
  {T\"orm\"a}}, \bibinfo {author} {\bibfnamefont {S.}~\bibnamefont {Peotta}},\
  and\ \bibinfo {author} {\bibfnamefont {B.~A.}\ \bibnamefont {Bernevig}},\
  }\bibfield  {title} {\bibinfo {title} {Superconductivity, superfluidity and
  quantum geometry in twisted multilayer systems},\ }\href
  {https://doi.org/10.1038/s42254-022-00466-y} {\bibfield  {journal} {\bibinfo
  {journal} {Nat. Rev. Phys.}\ }\textbf {\bibinfo {volume} {4}},\ \bibinfo
  {pages} {528} (\bibinfo {year} {2022})}\BibitemShut {NoStop}%
\bibitem [{\citenamefont {Iskin}(2018)}]{Iskin2018}%
  \BibitemOpen
  \bibfield  {author} {\bibinfo {author} {\bibfnamefont {M.}~\bibnamefont
  {Iskin}},\ }\bibfield  {title} {\bibinfo {title} {{Quantum-metric
  contribution to the pair mass in spin-orbit-coupled Fermi superfluids}},\
  }\href {https://doi.org/10.1103/PhysRevA.97.033625} {\bibfield  {journal}
  {\bibinfo  {journal} {Phys. Rev. A}\ }\textbf {\bibinfo {volume} {97}},\
  \bibinfo {pages} {033625} (\bibinfo {year} {2018})}\BibitemShut {NoStop}%
\bibitem [{\citenamefont {Gao}\ \emph {et~al.}(2023)\citenamefont {Gao} \emph
  {et~al.}}]{Gao2023NLH}%
  \BibitemOpen
  \bibfield  {author} {\bibinfo {author} {\bibfnamefont {A.}~\bibnamefont
  {Gao}} \emph {et~al.},\ }\bibfield  {title} {\bibinfo {title} {{Quantum
  metric nonlinear Hall effect in a topological antiferromagnetic
  heterostructure}},\ }\href {https://doi.org/10.1126/science.adf1506}
  {\bibfield  {journal} {\bibinfo  {journal} {Science}\ }\textbf {\bibinfo
  {volume} {381}},\ \bibinfo {pages} {181} (\bibinfo {year}
  {2023})}\BibitemShut {NoStop}%
\bibitem [{\citenamefont {Wang}\ \emph {et~al.}(2023)\citenamefont {Wang} \emph
  {et~al.}}]{Wang2023NLH}%
  \BibitemOpen
  \bibfield  {author} {\bibinfo {author} {\bibfnamefont {N.}~\bibnamefont
  {Wang}} \emph {et~al.},\ }\bibfield  {title} {\bibinfo {title}
  {Quantum-metric-induced nonlinear transport in a topological
  antiferromagnet},\ }\href {https://doi.org/10.1038/s41586-023-06363-3}
  {\bibfield  {journal} {\bibinfo  {journal} {Nature}\ }\textbf {\bibinfo
  {volume} {621}},\ \bibinfo {pages} {487} (\bibinfo {year}
  {2023})}\BibitemShut {NoStop}%
\bibitem [{\citenamefont {Ahn}\ \emph {et~al.}(2022)\citenamefont {Ahn},
  \citenamefont {Guo}, \citenamefont {Nagaosa},\ and\ \citenamefont
  {Vishwanath}}]{Ahn2022Riemannian}%
  \BibitemOpen
  \bibfield  {author} {\bibinfo {author} {\bibfnamefont {J.}~\bibnamefont
  {Ahn}}, \bibinfo {author} {\bibfnamefont {G.-Y.}\ \bibnamefont {Guo}},
  \bibinfo {author} {\bibfnamefont {N.}~\bibnamefont {Nagaosa}},\ and\ \bibinfo
  {author} {\bibfnamefont {A.}~\bibnamefont {Vishwanath}},\ }\bibfield  {title}
  {\bibinfo {title} {Riemannian geometry of resonant optical responses},\
  }\href {https://doi.org/10.1038/s41567-021-01465-z} {\bibfield  {journal}
  {\bibinfo  {journal} {Nat. Phys.}\ }\textbf {\bibinfo {volume} {18}},\
  \bibinfo {pages} {290} (\bibinfo {year} {2022})}\BibitemShut {NoStop}%
\bibitem [{\citenamefont {Yu}\ \emph {et~al.}(2024)\citenamefont {Yu},
  \citenamefont {Ciccarino}, \citenamefont {Bianco}, \citenamefont {Errea},
  \citenamefont {Narang},\ and\ \citenamefont {Bernevig}}]{Yu2024EPC}%
  \BibitemOpen
  \bibfield  {author} {\bibinfo {author} {\bibfnamefont {J.}~\bibnamefont
  {Yu}}, \bibinfo {author} {\bibfnamefont {C.~J.}\ \bibnamefont {Ciccarino}},
  \bibinfo {author} {\bibfnamefont {R.}~\bibnamefont {Bianco}}, \bibinfo
  {author} {\bibfnamefont {I.}~\bibnamefont {Errea}}, \bibinfo {author}
  {\bibfnamefont {P.}~\bibnamefont {Narang}},\ and\ \bibinfo {author}
  {\bibfnamefont {B.~A.}\ \bibnamefont {Bernevig}},\ }\bibfield  {title}
  {\bibinfo {title} {Non-trivial quantum geometry and the strength of
  electron--phonon coupling},\ }\href
  {https://doi.org/10.1038/s41567-024-02486-0} {\bibfield  {journal} {\bibinfo
  {journal} {Nat. Phys.}\ }\textbf {\bibinfo {volume} {20}},\ \bibinfo {pages}
  {1262} (\bibinfo {year} {2024})}\BibitemShut {NoStop}%
\bibitem [{\citenamefont {Kang}\ \emph {et~al.}(2025)\citenamefont {Kang},
  \citenamefont {Kim}, \citenamefont {Qian} \emph {et~al.}}]{Kang2024measure}%
  \BibitemOpen
  \bibfield  {author} {\bibinfo {author} {\bibfnamefont {M.}~\bibnamefont
  {Kang}}, \bibinfo {author} {\bibfnamefont {S.}~\bibnamefont {Kim}}, \bibinfo
  {author} {\bibfnamefont {Y.}~\bibnamefont {Qian}}, \emph {et~al.},\
  }\bibfield  {title} {\bibinfo {title} {Measurements of the quantum geometric
  tensor in solids},\ }\href {https://doi.org/10.1038/s41567-024-02678-8}
  {\bibfield  {journal} {\bibinfo  {journal} {Nature Physics}\ }\textbf
  {\bibinfo {volume} {21}},\ \bibinfo {pages} {110} (\bibinfo {year}
  {2025})}\BibitemShut {NoStop}%
\bibitem [{\citenamefont {Aharonov}\ and\ \citenamefont
  {Stern}(1992)}]{Aharonov1992}%
  \BibitemOpen
  \bibfield  {author} {\bibinfo {author} {\bibfnamefont {Y.}~\bibnamefont
  {Aharonov}}\ and\ \bibinfo {author} {\bibfnamefont {A.}~\bibnamefont
  {Stern}},\ }\bibfield  {title} {\bibinfo {title} {{Origin of the geometric
  forces accompanying Berry's geometric potentials}},\ }\href
  {https://doi.org/10.1103/PhysRevLett.69.3593} {\bibfield  {journal} {\bibinfo
   {journal} {Phys. Rev. Lett.}\ }\textbf {\bibinfo {volume} {69}},\ \bibinfo
  {pages} {3593} (\bibinfo {year} {1992})}\BibitemShut {NoStop}%
\bibitem [{\citenamefont {Hannay}(1985)}]{Hannay1985}%
  \BibitemOpen
  \bibfield  {author} {\bibinfo {author} {\bibfnamefont {J.~H.}\ \bibnamefont
  {Hannay}},\ }\bibfield  {title} {\bibinfo {title} {{Angle variable holonomy
  in adiabatic excursion of an integrable Hamiltonian}},\ }\href
  {https://doi.org/10.1088/0305-4470/18/2/011} {\bibfield  {journal} {\bibinfo
  {journal} {J. Phys. A: Math. Gen.}\ }\textbf {\bibinfo {volume} {18}},\
  \bibinfo {pages} {221} (\bibinfo {year} {1985})}\BibitemShut {NoStop}%
\bibitem [{\citenamefont {D’Alessio}\ and\ \citenamefont
  {Polkovnikov}(2014)}]{Dalessio2014}%
  \BibitemOpen
  \bibfield  {author} {\bibinfo {author} {\bibfnamefont {L.}~\bibnamefont
  {D’Alessio}}\ and\ \bibinfo {author} {\bibfnamefont {A.}~\bibnamefont
  {Polkovnikov}},\ }\bibfield  {title} {\bibinfo {title} {{Emergent Newtonian
  dynamics and the geometric origin of mass}},\ }\href
  {https://doi.org/https://doi.org/10.1016/j.aop.2014.03.009} {\bibfield
  {journal} {\bibinfo  {journal} {Ann. Phys.}\ }\textbf {\bibinfo {volume}
  {345}},\ \bibinfo {pages} {141} (\bibinfo {year} {2014})}\BibitemShut
  {NoStop}%
\bibitem [{\citenamefont {Kolodrubetz}\ \emph {et~al.}(2017)\citenamefont
  {Kolodrubetz}, \citenamefont {Sels}, \citenamefont {Mehta},\ and\
  \citenamefont {Polkovnikov}}]{Kolodrubetz2017}%
  \BibitemOpen
  \bibfield  {author} {\bibinfo {author} {\bibfnamefont {M.}~\bibnamefont
  {Kolodrubetz}}, \bibinfo {author} {\bibfnamefont {D.}~\bibnamefont {Sels}},
  \bibinfo {author} {\bibfnamefont {P.}~\bibnamefont {Mehta}},\ and\ \bibinfo
  {author} {\bibfnamefont {A.}~\bibnamefont {Polkovnikov}},\ }\bibfield
  {title} {\bibinfo {title} {Geometry and non-adiabatic response in quantum and
  classical systems},\ }\href {https://doi.org/10.1016/j.physrep.2017.07.001}
  {\bibfield  {journal} {\bibinfo  {journal} {Phys. Rep.}\ }\textbf {\bibinfo
  {volume} {697}},\ \bibinfo {pages} {1} (\bibinfo {year} {2017})}\BibitemShut
  {NoStop}%
\bibitem [{\citenamefont {Barrera}\ \emph {et~al.}(2026)\citenamefont
  {Barrera}, \citenamefont {Arovas}, \citenamefont {Chandran},\ and\
  \citenamefont {Polkovnikov}}]{Barrera2026}%
  \BibitemOpen
  \bibfield  {author} {\bibinfo {author} {\bibfnamefont {B.}~\bibnamefont
  {Barrera}}, \bibinfo {author} {\bibfnamefont {D.~P.}\ \bibnamefont {Arovas}},
  \bibinfo {author} {\bibfnamefont {A.}~\bibnamefont {Chandran}},\ and\
  \bibinfo {author} {\bibfnamefont {A.}~\bibnamefont {Polkovnikov}},\
  }\bibfield  {title} {\bibinfo {title} {{The moving Born–Oppenheimer
  approximation}},\ }\href {https://doi.org/10.1073/pnas.2507816123} {\bibfield
   {journal} {\bibinfo  {journal} {PNAS}\ }\textbf {\bibinfo {volume} {123}},\
  \bibinfo {pages} {e2507816123} (\bibinfo {year} {2026})}\BibitemShut
  {NoStop}%
\bibitem [{\citenamefont {Karplus}\ and\ \citenamefont
  {Luttinger}(1954)}]{Karplus1954}%
  \BibitemOpen
  \bibfield  {author} {\bibinfo {author} {\bibfnamefont {R.}~\bibnamefont
  {Karplus}}\ and\ \bibinfo {author} {\bibfnamefont {J.~M.}\ \bibnamefont
  {Luttinger}},\ }\bibfield  {title} {\bibinfo {title} {Hall effect in
  ferromagnetics},\ }\href {https://doi.org/10.1103/PhysRev.95.1154} {\bibfield
   {journal} {\bibinfo  {journal} {Phys. Rev.}\ }\textbf {\bibinfo {volume}
  {95}},\ \bibinfo {pages} {1154} (\bibinfo {year} {1954})}\BibitemShut
  {NoStop}%
\bibitem [{\citenamefont {Marzari}\ and\ \citenamefont
  {Vanderbilt}(1997)}]{Marzari1997}%
  \BibitemOpen
  \bibfield  {author} {\bibinfo {author} {\bibfnamefont {N.}~\bibnamefont
  {Marzari}}\ and\ \bibinfo {author} {\bibfnamefont {D.}~\bibnamefont
  {Vanderbilt}},\ }\bibfield  {title} {\bibinfo {title} {{Maximally localized
  generalized Wannier functions for composite energy bands}},\ }\href
  {https://doi.org/10.1103/PhysRevB.56.12847} {\bibfield  {journal} {\bibinfo
  {journal} {Phys. Rev. B}\ }\textbf {\bibinfo {volume} {56}},\ \bibinfo
  {pages} {12847} (\bibinfo {year} {1997})}\BibitemShut {NoStop}%
\bibitem [{\citenamefont {Marzari}\ \emph {et~al.}(2012)\citenamefont
  {Marzari}, \citenamefont {Mostofi}, \citenamefont {Yates}, \citenamefont
  {Souza},\ and\ \citenamefont {Vanderbilt}}]{Marzari2012}%
  \BibitemOpen
  \bibfield  {author} {\bibinfo {author} {\bibfnamefont {N.}~\bibnamefont
  {Marzari}}, \bibinfo {author} {\bibfnamefont {A.~A.}\ \bibnamefont
  {Mostofi}}, \bibinfo {author} {\bibfnamefont {J.~R.}\ \bibnamefont {Yates}},
  \bibinfo {author} {\bibfnamefont {I.}~\bibnamefont {Souza}},\ and\ \bibinfo
  {author} {\bibfnamefont {D.}~\bibnamefont {Vanderbilt}},\ }\bibfield  {title}
  {\bibinfo {title} {{Maximally localized Wannier functions: Theory and
  applications}},\ }\href {https://doi.org/10.1103/RevModPhys.84.1419}
  {\bibfield  {journal} {\bibinfo  {journal} {Rev. Mod. Phys.}\ }\textbf
  {\bibinfo {volume} {84}},\ \bibinfo {pages} {1419} (\bibinfo {year}
  {2012})}\BibitemShut {NoStop}%
\bibitem [{\citenamefont {Chang}\ and\ \citenamefont
  {Niu}(2008)}]{ChangNiu2008}%
  \BibitemOpen
  \bibfield  {author} {\bibinfo {author} {\bibfnamefont {M.-C.}\ \bibnamefont
  {Chang}}\ and\ \bibinfo {author} {\bibfnamefont {Q.}~\bibnamefont {Niu}},\
  }\bibfield  {title} {\bibinfo {title} {Berry curvature, orbital moment, and
  effective quantum theory of electrons in electromagnetic fields},\ }\href
  {https://doi.org/10.1088/0953-8984/20/19/193202} {\bibfield  {journal}
  {\bibinfo  {journal} {J. Phys.: Condens. Matter}\ }\textbf {\bibinfo {volume}
  {20}},\ \bibinfo {pages} {193202} (\bibinfo {year} {2008})}\BibitemShut
  {NoStop}%
\bibitem [{\citenamefont {Thonhauser}\ \emph {et~al.}(2005)\citenamefont
  {Thonhauser}, \citenamefont {Ceresoli}, \citenamefont {Vanderbilt},\ and\
  \citenamefont {Resta}}]{Thonhauser2005}%
  \BibitemOpen
  \bibfield  {author} {\bibinfo {author} {\bibfnamefont {T.}~\bibnamefont
  {Thonhauser}}, \bibinfo {author} {\bibfnamefont {D.}~\bibnamefont
  {Ceresoli}}, \bibinfo {author} {\bibfnamefont {D.}~\bibnamefont
  {Vanderbilt}},\ and\ \bibinfo {author} {\bibfnamefont {R.}~\bibnamefont
  {Resta}},\ }\bibfield  {title} {\bibinfo {title} {Orbital magnetization in
  periodic insulators},\ }\href {https://doi.org/10.1103/PhysRevLett.95.137205}
  {\bibfield  {journal} {\bibinfo  {journal} {Phys. Rev. Lett.}\ }\textbf
  {\bibinfo {volume} {95}},\ \bibinfo {pages} {137205} (\bibinfo {year}
  {2005})}\BibitemShut {NoStop}%
\bibitem [{\citenamefont {Resta}(2017)}]{Resta2017Drude}%
  \BibitemOpen
  \bibfield  {author} {\bibinfo {author} {\bibfnamefont {R.}~\bibnamefont
  {Resta}},\ }\bibfield  {title} {\bibinfo {title} {{Geometrical meaning of the
  Drude weight and its relationship to orbital magnetization}},\ }\bibfield
  {journal} {\bibinfo  {journal} {arXiv:1703.00712}\ }\href
  {https://doi.org/10.48550/arXiv.1703.00712} {10.48550/arXiv.1703.00712}
  (\bibinfo {year} {2017})\BibitemShut {NoStop}%
\bibitem [{\citenamefont {Shinada}\ and\ \citenamefont
  {Nagaosa}(2025)}]{Shinada2025bounds}%
  \BibitemOpen
  \bibfield  {author} {\bibinfo {author} {\bibfnamefont {K.}~\bibnamefont
  {Shinada}}\ and\ \bibinfo {author} {\bibfnamefont {N.}~\bibnamefont
  {Nagaosa}},\ }\bibfield  {title} {\bibinfo {title} {{Quantum geometric bounds
  for observables: Linear responses, Drude weight, and orbital
  magnetization}},\ }\href {https://doi.org/10.1103/qxbl-qd4f} {\bibfield
  {journal} {\bibinfo  {journal} {Phys. Rev. B}\ }\textbf {\bibinfo {volume}
  {112}},\ \bibinfo {pages} {155158} (\bibinfo {year} {2025})}\BibitemShut
  {NoStop}%
\bibitem [{\citenamefont {Ashcroft}\ and\ \citenamefont
  {Mermin}(1976)}]{ashcroft1976solid}%
  \BibitemOpen
  \bibfield  {author} {\bibinfo {author} {\bibfnamefont {N.~W.}\ \bibnamefont
  {Ashcroft}}\ and\ \bibinfo {author} {\bibfnamefont {N.~D.}\ \bibnamefont
  {Mermin}},\ }\href@noop {} {\emph {\bibinfo {title} {Solid State Physics}}}\
  (\bibinfo  {publisher} {Holt, Rinehart and Winston},\ \bibinfo {address} {New
  York},\ \bibinfo {year} {1976})\BibitemShut {NoStop}%
\end{thebibliography}

%

\appendix

\vspace{1em} 
\begin{center}
    {\large\bf End Matter}
\end{center}
\vspace{1em}

\textit{Adiabatic dynamics}.---We solve the equation of motion for $\boldsymbol{\ell}$ in Eq.\ \eqref{eq:eom} for $|\mathbf{B}(t)| = \mathrm{const}$. The rate of change $\dot{\mathbf{b}} = \boldsymbol{\omega} \times \mathbf{b}$ defines the component $\boldsymbol{\omega}_\perp = \mathbf{b} \times \dot{\mathbf{b}}$ of the angular velocity, which is perpendicular to $\mathbf{b}$. We introduce a rotation matrix $R(t)$ such that $\mathbf{b} = R(t) \mathbf{z}$ and define $\boldsymbol{\ell} = R\boldsymbol{\ell}'$. Taking a derivative of the first expression gives
\begin{equation}
    \mathbf{\dot b} = \frac{dR}{dt} R^\top \mathbf{b},
\end{equation}
so that we can write $\frac{dR}{dt}R^\top = \boldsymbol{\omega} \times$. For given basis vectors $\mathbf{e}_1$ and $\mathbf{e}_2$ in the plane perpendicular to $\mathbf{b}$, the component $\omega_\parallel$ parallel to $\mathbf{b}$ is fixed by the condition $\dot{\mathbf{e}}_j = \boldsymbol{\omega} \times \mathbf{e}_j$. Indeed, we observe that $\omega_\parallel = \mathbf{b}\cdot \boldsymbol{\omega} = \mathbf{e}_2\cdot (\boldsymbol{\omega} \times \mathbf{e}_1) = 
\mathbf{e}_2\cdot \dot{\mathbf{e}}_1$. 

The equation of motion for $\boldsymbol{\ell}$ implies
\begin{equation}
    R \frac{d\boldsymbol{\ell}'}{dt} + \frac{dR}{dt} \boldsymbol{\ell}' = B R(\boldsymbol{\ell}' \times \mathbf{z}). 
\end{equation}
Multiplying from the left by $R^\top$ gives
\begin{equation}
    \frac{d\boldsymbol{\ell}'}{dt} = B \boldsymbol{\ell}' \times \mathbf{z} - R^\top \frac{dR}{dt} \boldsymbol{\ell}'.
\end{equation}
We define $\boldsymbol{\omega}'$ through $R^\top \frac{dR}{dt} = \boldsymbol{\omega}' \times$ and identify $\boldsymbol{\omega}'$ as the angular velocity $\boldsymbol{\omega}$ rotated into the moving coordinate frame, $\boldsymbol{\omega} = R \boldsymbol{\omega}'$. To see this, we note that $\boldsymbol{\omega} \times = R (\boldsymbol{\omega}' \times) R^\top$. We then find the equation of motion
\begin{equation}
    \frac{d\boldsymbol{\ell}'}{dt}
     = \boldsymbol{\ell}' \times ( B\mathbf{z} + \boldsymbol{\omega}' ) 
     \label{eq:eomcomoving}
\end{equation}
in the moving coordinate system. 

Solving Eq.\ \eqref{eq:eomcomoving} and rotating the solution back to the original coordinate frame, we find the solution
\begin{equation}
    \boldsymbol{\ell} \simeq \ell_\parallel (\mathbf{b} + \frac{1}{B} \mathbf{b} \times \dot{\mathbf{b}}) + \ell_\perp [\mathbf{e}_1 \cos\Phi(t)  - \mathbf{e}_2 \sin\Phi(t)]
\end{equation}
with $\dot\Phi = B + \omega_\parallel$ and 
\begin{equation}
  \Phi(t) = Bt + \int_0^t dt' \mathbf{e}_2(t')\cdot \dot{\mathbf{e}}_1(t') + \Phi_0.
\end{equation}
The phase offset $\Phi_0$ is fixed by the initial conditions. If $\mathbf{e}_1$ and $\mathbf{e}_2$ are parallel transported along the loop of $\mathbf{b}$, they satisfy $\mathbf{e}_2\cdot \dot{\mathbf{e}}_1 = 0$ and hence $\Phi(t) = Bt + \Phi_0$, but the coordinate frame does not return to itself at the end of the closed loop. Equations \eqref{eq:Bprime} and \eqref{eq:oscvel} are written with this assumption.

Alternatively, we can choose $\mathbf{e}_1 = \dot{\mathbf{b}}$ and $\mathbf{e}_2 =(\mathbf{b} \times \dot{\mathbf{b}})/|\dot{\mathbf{b}}|$. This coordinate frame returns to itself after a closed loop of $\mathbf{b}$ and Hannay's angle $\int_0^T dt' \mathbf{e}_2(t')\cdot \dot{\mathbf{e}}_1(t')$ can be expressed in terms of the enclosed solid angle. 

\textit{Velocity and kinetic energy for a particle pair with arbitrary interaction potential}.---We consider the velocity of particle $a$ in one dimension, averaged over the fast oscillations. It takes the form  
\begin{equation}
  \langle {\dot x}_a \rangle = - B\lperp
  \frac{\partial^2\bhat}{\partial{p}_a^2} \cdot \langle \left({\mathbf{e}}_{a,1}\cos Bt - {\mathbf{e}}_{a,2}\sin Bt\right) \delta {p}_a \rangle .
\end{equation}
Under the time average we can replace $\delta p_a$ by $p_a$. Integrating by parts gives
\begin{equation}
  \langle{\dot x}_a  \rangle= \lperp \frac{\partial^2\bhat}{\partial{p}_a^2} \cdot \langle
  \left({\mathbf{e}}_{a,1}\sin Bt + {\mathbf{e}}_{a,2}\cos Bt\right) {\dot p}_a \rangle.
\end{equation}
Using that the momenta $p_a$ obey ${\dot p}_a = -\partial_a V({x}_1-{ x}_2)$, we obtain
\begin{equation}
  \langle {\dot x}_a \rangle = - \lperp \frac{\partial^2\mathbf{b}}{\partial{p}_a^2}\cdot \langle
  \left({\mathbf{e}}_{a,1}\sin Bt + {\mathbf{e}}_{a,2}\cos Bt\right) \partial_a V(x_1-x_2) \rangle ,
\end{equation}
Fourier transforming
\begin{equation}
    V(x_1 - x_2) = \int \frac{dq}{2\pi} V_q e^{iq(x_1 - x_2)},
\end{equation}
and decomposing the positions of the particles into slowly and rapidly varying contributions, $x_a = \langle x_a \rangle + \zeta_a$,  we find
\begin{align}
  \langle {\dot x}_a \rangle & = - \lperp \int \frac{dq}{2\pi} V_q \partial_a e^{iq(\langle x_1 \rangle - \langle x_2\rangle)} \,
\frac{\partial^2\bhat}{\partial{p}_a^2} 
\nonumber\\
& \qquad \cdot \langle \left({\mathbf{e}}_{a,1}\sin Bt + {\mathbf{e}}_{a,2}\cos Bt\right) e^{iq(\zeta_1 - \zeta_2)} \rangle .
\end{align}
We assume that the precessions of the two particles dephase sufficiently rapidly, so that we can factorize the average. This yields 
\begin{align}
  \langle {\dot x}_1 \rangle & = - \lperp \int \frac{dq}{2\pi} iq V_q  e^{iq(\langle x_1 \rangle - \langle x_2\rangle)}
\frac{\partial^2\bhat}{\partial{p}_1^2} 
\nonumber\\
& \qquad \cdot \langle \left({\mathbf{e}}_{1,1}\sin Bt + {\mathbf{e}}_{1,2}\cos Bt\right) e^{iq \zeta_1} \rangle \langle e^{- iq \zeta_2} \rangle 
\end{align}
for $a=1$, and a corresponding result for $a=2$. To compute the averages, we note that from Eq.\ \eqref{eq:oscvel}, 
\begin{equation}
    \zeta_a = - \lperp
  \left({\mathbf{e}}_{a,1}\sin Bt + {\mathbf{e}}_{a,2}\cos Bt\right)\cdot
  \frac{\partial\bhat}{\partial{p}_a} .
\end{equation}
This gives 
\begin{equation}
    \langle  e^{- iq\zeta_2)} \rangle = J_0\left(q\ell_\perp \left|\frac{\partial\bhat}{\partial{p}_2}\right|\right)
\end{equation}
and 
\begin{align}
   & \langle \left({\mathbf{e}}_{1,1}\sin Bt + {\mathbf{e}}_{1,2}\cos Bt\right) e^{iq\zeta_1 } \rangle 
   \nonumber\\
   & \qquad\qquad\qquad = -i \widehat{\frac{\partial\bhat}{\partial{p}_1}} J_1 \left(q\ell_\perp \left|\frac{\partial\bhat}{\partial{p}_1}\right|\right).
\end{align}
Here, the hat denotes the corresponding unit vector. Inserting these averages into the expression for the velocity yields (from here on,  we drop the angular brackets for the time average of $x_1,x_2$ unless needed to avoid confusion), 
\begin{align}
  {\dot x}_1 &=  i \lperp \int \frac{dq}{2\pi} i q  V_q e^{iq( x_1  -  x_2)}
  \frac{\partial^2\bhat}
  {\partial{p}_1^2} \cdot \widehat{\frac{\partial\bhat}{\partial{p}_1}} 
  \nonumber\\
  & \qquad \times J_1 \left(q\ell_\perp \left|\frac{\partial\bhat}{\partial{p}_1}\right|\right)
  J_0\left(q\ell_\perp \left|\frac{\partial\bhat}{\partial{p}_2}\right|\right).
\end{align}
We observe that this can be rewritten as 
\begin{align}
   {\dot x}_1 &=   \partial_{p_1}
  \int \frac{dq}{2\pi}  V_q e^{iq(x_1 -  x_2 )} 
  \nonumber\\
  & \qquad \times J_0 \left(q\ell_\perp \left|\frac{\partial\bhat}{\partial{p}_1}\right|\right)
  J_0\left(q\ell_\perp \left|\frac{\partial\bhat}{\partial{p}_2}\right|\right).
  \label{eq}
\end{align}
As a result of the average, the Fourier components of the interaction are modified by Bessel-function form factors. Using Eq.\ \eqref{eq:xi}, we can identify the arguments of the Bessel functions with $q\xi_a$. We note that the expression in Eq.\ \eqref{eq:twoparticle} for a harmonic potential, applied to a one-dimensional system, is reproduced by expanding the Bessel functions to leading nonvanishing order in $q$. 

We can view this velocity as originating from an emergent kinetic energy
\begin{equation}
   H_\mathrm{kin} =  \int \frac{dq}{2\pi}  V_q e^{iq(x_1 - x_2 )} 
 J_0 (q\ell_\perp \xi_1)
  J_0 (q\ell_\perp \xi_2)
\end{equation}
of the particle pair. This is just the interaction potential of the pair averaged over the fast oscillations of the two particles, $H_\mathrm{kin} = \langle V(x_1 - x_2)\rangle$. The dependence on the particle momenta enters through the displacement amplitudes $\xi_a$. Importantly, in addition to the momenta, the kinetic energy also depends on the average distance between the particles $x_1-x_2$. 
\end{document}